# High-temperature sensing using a hollow-core fiber with thick cladding tubes

Gabriel Labes Rodrigues, Cristiano M. B. Cordeiro, Foued Amrani, Frédéric Gérôme, Fetah Benabid and Jonas H. Osório

*Abstract*—We report on high-temperature sensing measurements using a tubular-lattice hollow-core photonic crystal fiber displaying a microstructure formed of eight 2.4 µm-thick cladding tubes. The larger thickness of our fiber's cladding tubes compared to other hollow fibers operating in the visible and infrared ranges entails multiple narrow transmission bands in its transmission spectrum (6 bands in the spectral range between 400 nm and 950 nm) and benefits the realization of the temperature sensing measurements. The principle of operation of our device is based on the thermo-optic effect and thermal expansion-induced spectral shifts of the fiber transmission bands due to temperature variations. To study the sensor operation, we monitored the fiber transmission bands' spectral positions from room temperature to 1085 ºC in both ramp-up and ramp-down scenarios. Additionally, we investigated the optimization opportunities by assessing an analytical model describing the fiber transmission characteristics and discussed the alternatives for enhancing the sensor performance. Moreover, our fiber characterization experiments revealed a consistent confinement loss trend aligned with the scaling laws in tubular-lattice hollow-core fibers. We thus understand that the results presented in this manuscript highlight a relevant path for the development of temperature sensors based on microstructured hollow-core optical fibers endowed with thick cladding tubes.

## I. Introduction

Optical fibers are undoubtedly excellent platforms for the development of sensing systems. By relying on different technologies, such as fiber gratings, post-processed or specialty fibers, to mention a few, a myriad of physical, chemical, and biological fiber-based sensors could be demonstrated up to date. Remarkably, these setups can be set in very compact configurations and work in harsh environments (*e.g.*, corrosive, explosive, or subjected to high temperatures), hence figuring as very promising application alternatives to situations in which standard electrical sensors have limited performance or cannot be deployed.

Thus, in the framework of harsh environment sensing, one can identify the development of fiber-based high-temperature sensors as an active field of research, especially due to their potential applications in areas such as aviation, metallurgy, and fossil fuel production [1]. High-temperature sensors based on fiber Bragg gratings [2], Fabry-Perot interferometers [3], fiber tapers [4], multicore, suspended-core, and capillary fibers [5-7] have, therefore, been investigated. Additionally, one notes significant effort by the optical fiber sensing community

The authors thank FAPESP (grants 2022/08805-7 and 2021/13097-9), CNPq (grants 309989/2021-3 and 305024/2023-0), and FAPEMIG (grant RED-00046-23) for financial support.
G. L. R. and C. M. B. C. are with the Institute of Physics Gleb Wataghin, University of Campinas, Campinas, Brazil (email: g251066@dac.unicamp.br and cordeiro@unicamp.br).
F.A., F. G., and F. B. are with the XLIM Institute, University of Limoges, Limoges, France (email: fouedamrani@glophotonics.fr, frederic.gerome@xlim.fr, f.benabid@xlim.fr).
J. H. O. was with the Institute of Physics Gleb Wataghin, University of Campinas, Campinas, Brazil. He is now with the Department of Physics, Federal University of Lavras, Lavras, Brazil (email: jonas.osorio@ufla.br).

towards the development of sapphire fiber sensors, particularly regarding applications in which temperatures above 1000 ºC are required [8].

In a broader scenario, a particular family of optical fibers displaying an empty core and a microstructured cladding, namely hollow-core photonic crystal fibers (HCPCF) [9], emerges as an enabling technology for addressing sensing challenges. Thanks to their alveolar microstructures, these fibers can be gas- and liquid-filled, hence disclosing a wide set of opportunities regarding, for example, the detection of gas species [10] and the monitoring of chemical reactions [11]. Additionally, a judicious choice of the HCPCF microstructure parameters enables the application of these fibers as curvature and displacement sensors [12, 13].

Hence, given the interest in the development of high-temperature sensors and the current availability of high-performance HCPCFs [14], in this investigation, we explore the utilization of a single-ring tubular-lattice (SR-TL) HCPCF as a high-temperature sensing platform. The fiber we employ in our experiments is a SR-TL HCPCF with a microstructure formed of a set of eight 2.4 μm-thick cladding tubes that define the fiber's hollow core. The use of a SR-TL HCPCF endowed with such thick cladding tubes allowed us to attain, in the fiber transmission spectrum, several narrow transmission bands whose spectral positions could be readily followed while the external environment temperature was altered in the temperature sensing measurements.

Our characterization experiments allowed us to demonstrate a temperature sensor working from room temperature to 1085 ºC. Additionally, benefiting from the existence of 6 transmission bands within the wavelength span from 400 nm to

950 nm, we could recognize, while characterizing the fiber loss spectrum, the fiber loss trend predicted by previously reported scaling laws in SR-TL HCPCF [15]. Moreover, by assessing an analytical model describing the fiber transmission characteristics, we could analyze the sensor optimization opportunities. Thus, we believe that the results and analyses reported in this manuscript enlarge the framework of HCPCF applications and provide a valid alternative for the realization of high-temperature sensing measurements.

## II. Fiber characterization

The cross-section of the fiber we use in this investigation is presented in Fig. 1a. It stands for a SR-TL HCPCF whose microstructure encompasses a set of eight 2.4 μm-thick tubes that defines a hollow core with a 25 μm diameter. The fiber, which has been fabricated by following typical stack-and-draw techniques, has an outer diameter of 320 μm. Fig. 1b shows the transmission spectra referring to 47 m and 3 m-long fiber pieces, measured by using a supercontinuum light source and an optical spectrum analyzer during a cutback measurement routine.

Whilst our fiber displays thicker tubes compared to other SR-TL HCPCFs with operation in the visible and infrared ranges reported in the literature, one can recognize six transmission bands within the wavelength spanning from 450 nm to 950 nm. While having several narrow transmission bands is typically undesirable in most typical HCPCF applications due to bandwidth limitations, this property is of particular importance to our temperature sensing demonstration, which relies on tracking the fiber transmission bands' spectral positions as one alters the temperature the fiber experiences. We delve into the sensing platform details in the next section of this manuscript.

The fiber transmission spectra accounted from the cutback measurement allow determining the fiber loss spectrum shown in Fig. 1c. In the studied wavelength range, the minimum loss value has been measured as 14.5 dB/km at 484 nm. Additionally, as expected, we observe that the measured loss values increase for larger wavelengths. Indeed, this agrees with the confinement loss (CL) scaling laws in SR-TL HCPCFs [15]. Specifically, such scaling laws state that the CL minimum within every transmission band increases for larger wavelengths following a $\lambda^{4.5}$ trend, where $\lambda$ is the wavelength [15].

Regarding the latter, we present in Fig. 1d a graph of the measured loss minima as a function of the wavelength (blue squares) and the curve resulting from a nonlinear fitting using $CL = A \lambda^{4.5}$, where $A$ is a fitting constant. Remarkably, we observe that the measured loss minimum within each transmission band is consistent with the CL trend predicted by the scaling laws in SR-TL HCPCFs. In our analysis, the fitting constant $A$ has been accounted as $(1.37 \pm 0.02)$ dB/km.nm. We mention that, in the analyses presented above, the other sources of loss in HCPCFs (such as scattering and microbending loss [16]) have been disregarded. This is sound because the fiber we study herein displays greater confinement loss values than other, ultralow-loss, HCPCF reported in the literature [14]. Hence, its loss spectrum is still dominated by the fiber CL.

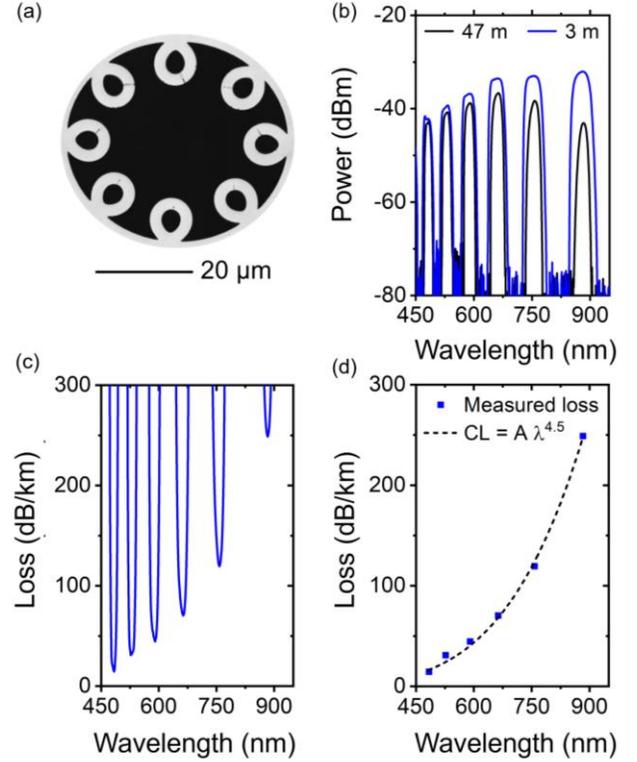

Fig. 1. (a) Fiber cross-section. (b) Transmission spectra of fiber pieces with lengths 47 m and 3 m accounted during a cutback measurement. (c) Measured loss spectrum between 450 nm and 950 nm. (d) Plot of the measured loss minima as a function of the wavelength (blue squares) and the corresponding fitted CL trend (dashed curve).

## III. Analytical study of the temperature sensing platform

The characteristic spectrum of SR-TL HCPCFs alternates regions of high- and low-loss which defines its transmission bands. The spectral positions of the latter are determined by the refractive indices of the core and cladding elements $n_{core}$ and $n_{clad}$, respectively, and by its struts' thicknesses $t$. The high-loss wavelengths, $\lambda_m$, can be approximately found by Eq. (1), where $m$ is a natural number, and referred to as the resonance order [9]. Observation of Eq. (1) allows recognizing that if the refractive index or the thickness of the cladding elements is altered, spectral shifting of the fiber transmission bands is expected. In our sensing platform, the spectral shifting of the fiber transmission bands is mediated by the thermo-optic effect and by the thermal expansion of the cladding elements, as these effects imply variation of $n_{clad}$ and $t$ values.

$$\lambda_m = \frac{2t}{m}\sqrt{n_{clad}^2 - n_{core}^2} \qquad (1)$$

To analyze the applicability of our SR-TL HCPCF as a temperature sensor, we apply an analytical model for calculating the complex effective refractive index of the leaky modes transmitted through capillary fibers as a function of the wavelength [17]. Although this model is simplified with respect to the fiber used in our experiments, it allows us to reproduce the main characteristics of the fiber transmission spectrum.

Thus, by investigating the simulated spectra, we can qualitatively analyze the main parameters of our sensing platform, namely its principle of operation, sensitivity, and resolution.

Assuming that light propagates in the fundamental mode, $HE_{11}$, we can use Eq. (2) to calculate its loss coefficient in units of m$^{-1}$, $\alpha$. In Eq. (2), $\phi = k_0 t\sqrt{n_{clad}^2 - n_{core}^2}$, $k_0 = 2\pi/\lambda$, and $t$, $n_{clad}$, and $n_{core}$ are defined as in Eq. 1 [17]. In turn, $\epsilon = (n_{clad}/n_{core})^2$, $j_{01} \approx 2.405$ is the first zero of the zero-order Bessel function, and $R$ is the core radius [17].

$$\alpha = \left(\frac{1+\cot^2\phi}{\epsilon-1}\right)\left(\frac{j_{01}^3}{k_0^3 n_{core}^3 R^4}\right)\left(\frac{\epsilon^2+1}{2}\right) \quad (2)$$

Thus, we can account for the fiber normalized transmission spectrum using $P[dB] = 10\log_{10}[e^{-\alpha L}]$, where $L$ is the length of the fiber. Fig. 2a shows the resulting spectrum calculated for an air-filled fiber with a length of 3 cm and a core radius of 12.5 µm between 450 nm and 950 nm. We comment that we chose the length of 3 cm because it was the one used in our experiments, to be described in the following.

Considering the latter, we explored the thermo-optic effect and thermal expansion impact on the fiber transmission spectrum by monitoring the spectral shifts of the fiber transmission bands as a function of temperature variations. Here, we use $dn_{clad}/dT = 1.090 \times 10^{-5}$ K$^{-1}$ as the thermo-optic coefficient and $\gamma = 5 \times 10^{-7}$ K$^{-1}$ as the thermal expansion coefficient of the glass, respectively [18, 19]. Inspection of Eq. (1) readily informs that both thermo-optic and thermal expansion contributions entail shifting of the transmission bands towards longer wavelengths as the temperature is increased.

Fig. 2b shows the calculated transmission spectrum between 540 nm and 585 nm for different temperature variations, $\Delta T$, from $\Delta T = 0$ ºC (corresponding to room temperature) to $\Delta T = 1000$ ºC. Observation of Fig. 2b allows recognizing that the fiber transmission dip redshifts as the temperature is increased, as expected. In the simulations shown in Fig. 2, for simplicity, we have set the refractive index of glass at room temperature as $n_{clad} = 1.45$. The corresponding refractive index values at different temperatures, in turn, has been accounted as $n_{clad}(\Delta T) = n_{clad} + (dn_{clad}/dT)\Delta T$. The refractive index of air has been set to $n_{core} = 1$.

We remark that the transmission minima of the simulated spectra are not visible in Fig. 2 because the corresponding points diverge when estimating the transmission spectrum in decibels. Nevertheless, if we define the position of the transmission minimum as the average of the wavelengths whose calculated transmitted power is lower than -60 dB, one can account for the wavelength shift of the transmission dip as a function of the temperature variation and determine a sensitivity of approximately 8 pm/ºC, which is slightly lower than the ones typically measured with standard fiber Bragg gratings sensors, generally around 10 pm/ºC [20]. Therefore, we can observe that the fiber platform we explore herein can be employed in temperature sensing tests, as we will detail in the following sections.

Moreover, further inspection of Eq. (1) allows us to observe that the shifting of the resonances is dominated by the thermo-optic effect, although the contribution of the thermal expansion cannot be neglected. Considering the parameters corresponding to the representative case shown in Fig. 2b and considering separately the thermo-optic effect and thermal expansion impact on the spectral position of the resonances, we can estimate that the thermo-optic effect contribution to the resonance shifting is about 25 times larger than that of the thermal expansion.

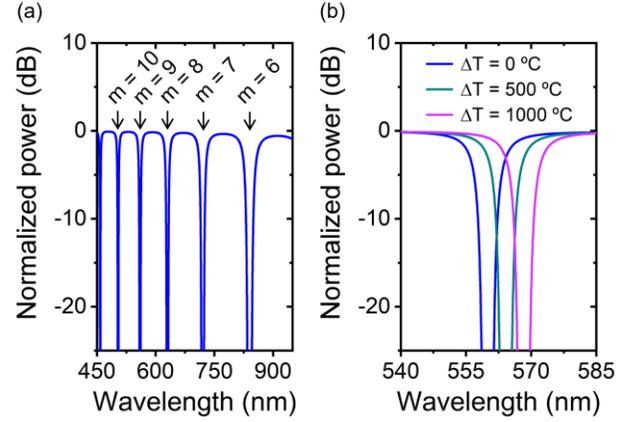

Fig. 2. Transmission spectra calculated analytically. (a) Graph of the calculated fiber transmission spectrum for an air-filled fiber with a length of 3 cm from 450 nm to 950 nm. The resonance orders (m) have been identified. (b) Calculated transmission spectrum around a representative transmission dip as a function of the temperature variation, $\Delta T$, in the spectral range between 540 nm to 585 nm.

## IV. HIGH-TEMPERATURE SENSING RESULTS

Fig. 3a shows the experimental setup used in this investigation, which employs a supercontinuum light source (SC) and an optical spectrum analyzer (OSA). The sensing platform is built up by splicing a 3 cm-long piece of the previously mentioned HCPCF in between singlemode and multimode fibers (SMF and MMF, respectively). The SMF is used to launch light into the HCPCF and the MMF to collect the transmitted light. We mention that the splices have been performed by using an arc fusion splicer working in manual mode; the arc parameters, namely its exposure time and intensity, have been adjusted to attain a robust fusion point between the fibers and suitable optical coupling between the fibers in terms of transmitted power, although the splice loss has not been optimized. In turn, to heat the fiber during the temperature sensing tests, we used a micro-heater (MH), indicated as a dashed rectangle in Fig 3a. We remark that the fiber length of 3 cm has been chosen to ensure that the full HCPCF section could be accommodated into our MH housing.

Fig. 3b presents the measured transmission spectra between 450 nm and 950 nm corresponding to the SMF-HCPCF-MMF system at representative temperatures, from room conditions to 1085 ºC. As predicted by the theoretical model, the transmission bands are redshifted as the temperature increases. A zoom in the spectral region corresponding to the resonances with orders m = 8, 9, 10 (transmission minima) is presented in Fig. 3c for better visualization of the wavelength shifts due to the temperature variations.

Fig. 4 displays the wavelength shift of the resonances shown in Fig. 4b for different temperature levels between 25 °C and 1085 °C in temperature ramp-up and ramp-down experiments. Here, we defined the spectral position of the resonances as the minimum obtained by fitting a second-degree polynomial for points with power levels lower than -40 dBm. The dashed curves in Fig. 4 indicate the lines fitted to the experimental data which, in turn, allows us to estimate our device's sensitivity. Table I exhibits the sensitivity values attained for the resonances with orders m = 8, 9, 10 during the temperature ramp-up and ramp-down tests.

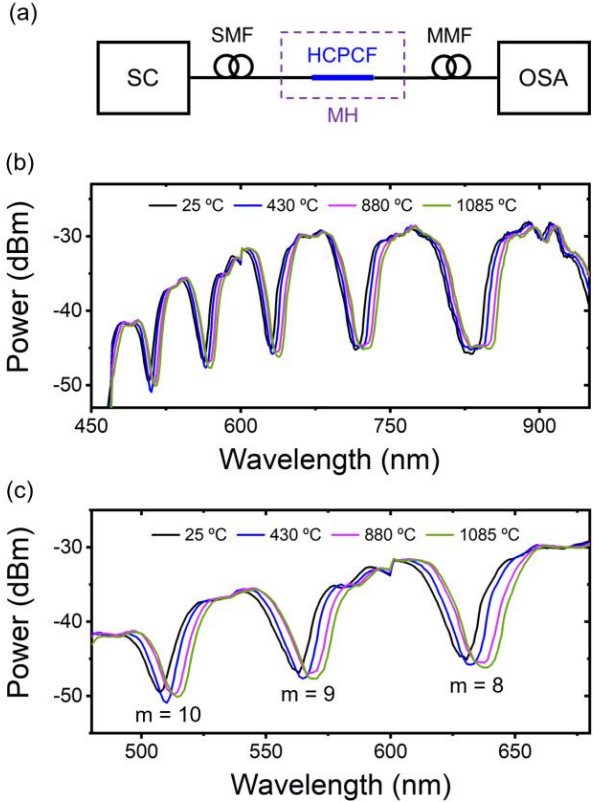

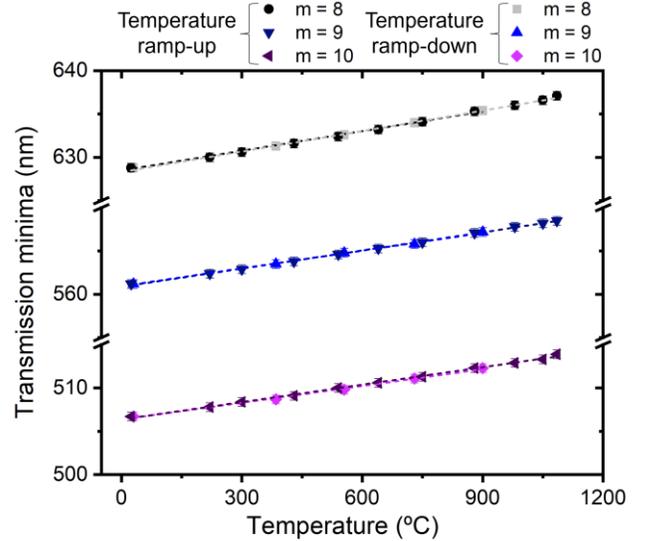

Fig. 3. (a) Diagram of the experimental setup. SC: supercontinuum source; OSA: optical spectrum analyzer; HCPCF: hollow-core photonic crystal fiber; SMF: singlemode fiber; MMF: multimode fiber; MH: microheater. (b) Measured transmission spectra between 450 nm and 950 nm at representative temperature levels and (c) zoom in the spectral region between 480 nm and 680 nm for better visualization of the transmission minima spectral shift due to temperature variations. The resonances with orders m = 8, 9, 10 have been identified.

Fig. 4. Spectral positions of the transmission minima with orders m = 8, 9, 10 during temperature ramp-up and ramp-down experiments. The dashed lines correspond to the curves fitted to the experimental data. The adjusted angular coefficients (sensitivities) are shown in Table I.

TABLE I
EXPERIMENTALLY DETERMINED SENSITIVITY VALUES

| Resonance order | Temperature test | Sensitivity (pm/°C) |
| --- | --- | --- |
| m = 8 | Ramp-up | (7.8 ± 0.1) |
|  | Ramp-down | (7.6 ± 0.2) |
| m = 9 | Ramp-up | (7.0 ± 0.2) |
|  | Ramp-down | (6.8 ± 0.3) |
| m = 10 | Ramp-up | (6.7 ± 0.1) |
|  | Ramp-down | (6.4 ± 0.1) |

Observation of the results shown in Table I allows recognizing that the sensitivities obtained during the temperature ramp-up and ramp-down experiments are consistent within our error bars. Additionally, we note that our results allow observing the sensitivity decreasing trend with increasing resonance order, as expected by Eq. (1). Moreover, if we consider that, in our measurements, we can resolve a wavelength shift of 0.2 nm, we can estimate our best resolution level as approximately 25 °C, corresponding to the results attained with the 8$^{th}$-order resonance.

## V. OPTIMIZATION OPPORTUNITIES

In the last section, we presented the characterization of our sensing platform and determined its corresponding sensitivity and resolution levels. To investigate our sensor's optimization possibilities, we used Eq. (2) to account for the transmission spectra of fibers with different strut thicknesses. Subsequently, we accounted for the expected sensitivity and resonances' full-width-at-half-maximum (FWHM) corresponding to each studied case. Thus, similarly to the analysis reported in [21] for a different sensing structure, we accounted for the ratio S/FWHM taking it as a metric to analyze the sensor's performance enhancement.

In our study, we considered fibers with struts thicknesses of t = 1 μm, t = 2.4 μm, and t = 3 μm, hence taking into account fibers with tube thicknesses smaller and larger than that of the fiber we used in our experiments (2.4 μm). The results of our theoretical analysis are shown in Fig. 5, where we plotted the expected sensitivity (S, Fig. 5a), resonances FWHM (Fig. 5b), and the ratio S/FWHM (Fig. 5c) for the fibers with different strut thicknesses (here considering the resonances lying within the spectral range between 450 nm and 950 nm; the resonances have been identified by their orders, m). We remark that, as can be expected by Eq. (1), the number of resonances in the considered spectral range for the fiber with t = 1 μm is much lower than that corresponding to the fiber with t = 3 μm (two resonances for the fiber with t = 1 μm and eight resonances for the fiber with t = 3 μm).

By observing Fig. 5a, we note that, as could be anticipated by Eq. (1), using resonances with lower orders (smaller m) entails larger sensitivities. However, the use of lower-order resonances also implies larger resonances FWHM, as can be seen in Fig. 5b. Indeed, the FWHM values variation can be very significant, particularly for fibers with larger strut thicknesses. For the fiber with t = 3 μm, for example, an approximately ten-fold reduction is observed between the FWHM values corresponding to the resonances with orders m = 7 and m = 14. This readily impacts the sensor resolution, as having narrower resonances allows for resolving smaller wavelength shifts.

As mentioned before, we plot in Fig. 5c the ratio S/FWHM as a metric for analyzing the sensor performance. As, in general, higher sensitivities and lower FWHM are desired, the S/FWHM ratio should be maximized for an enhanced sensing operation.

Due to the significative resonances' FWHM reduction for higher-order resonances in fibers with larger tube thickness, the fiber with t = 3 μm is the one that provides a better S/FWHM ratio. In particular, for its 14$^{th}$-order resonance, we observe a 2.5-fold augmentation of the S/FWHM ratio compared to the parameters corresponding to the fiber used in our experiments (*i.e.*, t = 2.4 μm and an 8$^{th}$-order resonance for our best experimental sensitivity). This indicates that, more generally, achieving optimal conditions for high-temperature sensing with SR-TL HCPCFs involves using fibers with the largest tube thickness possible and tracking resonances with the highest orders in the detectable spectral range.

Furthermore, we mention that, in a broader temperature sensing scenario also considering cases in which lower temperature monitoring is aimed, additional opportunities could be considered. In this context, we mention the opportunity of filling the SR-TL HCPCF microstructure with liquids [22]. This procedure allows, by exploring the larger thermo-optic coefficients of liquids comparatively with air and silica, to attain suitable temperature sensitivity for detecting lower temperature variations. Additionally, we can indicate the possibility of increasing the cladding tube thicknesses via coating [23]. Besides positively impacting the S/FWHM values by the tube thickness enlargement (as studied in Fig. 5), this strategy could potentially bring additional benefits in terms of sensitivity if the coating material has larger thermo-optic and thermal expansion coefficients compared to that of silica. Finally, we can also indicate the possibility of using all-polymer tubular HCPCFs [24], which could also be beneficial for temperature sensing due to the greater thermo-optic and thermal expansion coefficients of polymeric materials compared to that of glass. This opportunity, however, would be restricted to low-temperature sensing applications.

## VI. Conclusion

In this manuscript, we investigated the application of a tubular-lattice HCPCF exhibiting thick-cladding tubes as a high-temperature sensing platform. By assessing an analytical model describing the fiber transmission characteristics, we could describe the sensor's principle of operation, which is based on tracking the spectral shifts of the fiber transmission bands, mediated by the thermo-optic effect and thermal expansion, due to temperature variations. Our experiments allowed us to characterize the sensor response in both temperature ramp-up and ramp-down scenarios from room temperature and 1085 ºC. Additionally, we studied the opportunities for enhancing the sensor performance by investigating fibers with different cladding tube thicknesses and resonances with distinct orders. Furthermore, we discussed additional possibilities regarding the use of liquid-filled, coated, or polymer-made tubular fibers in a wider temperature-sensing framework. We understand that our investigation and corresponding analyses contribute to enlarging the context of HCPCF sensing applications and providing an additional alternative for the realization of high-temperature sensing besides traditional optical fiber platforms such as fiber gratings and interferometers.

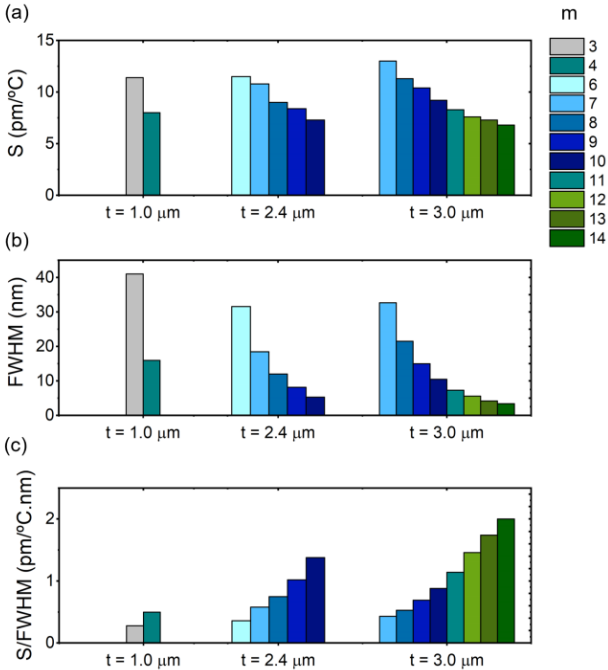

Fig. 5. Optimization study. Analytically calculated (a) sensitivity, S, (b) FWHM, and (c) the S/FWHM ratio for resonances with different orders (m) lying in the spectral range between 450 nm and 950 nm considering fibers with different tube thicknesses (t).